\documentclass{article}
\usepackage{cite}
\usepackage{graphicx}
\usepackage{dcolumn}

\begin{document}

\date{}
\title{On the application of the Lindstedt-Poincar\'{e} method to the
Lotka-Volterra system}
\author{Paolo Amore\thanks{%
e--mail: paolo.amore@gmail.com} \\
Facultad de Ciencias, CUICBAS, Universidad de Colima,\\
Bernal D\'{i}az del Castillo 340, Colima, Colima,Mexico \\
and \\
Francisco M. Fern\'andez\thanks{%
e--mail: framfer@gmail.com} \\
INIFTA (CONICET), Divisi\'{o}n Qu\'{i}mica Te\'{o}rica,\\
Blvd. 113 y 64 (S/N), Sucursal 4, Casilla de Correo 16,\\
1900 La Plata, Argentina}
\maketitle

\begin{abstract}
We apply the Lindstedt-Poincar\'{e} method to the Lotka-Volterra
model and discuss alternative implementations of the approach. By
means of an efficient systematic algorithm we obtain an
unprecedented number of perturbation corrections for the two
dynamical variables and the frequency. They enable us to estimate
the radius of convergence of the perturbation series for the
frequency as a function of the only model parameter. The method is
suitable for the treatment of systems with any number of dynamical
variables.
\end{abstract}

\section{Introduction}

\label{sec:intro}

In the last three decades there has been some interest in the application of
perturbation theory to nonlinear dynamical systems such as the
Lotka-Volterra model. Murty et al\cite{MSP90} applied perturbation theory to
a three-species ecological system and obtained the first perturbation
correction to the population of each species. However, they did not take
into account that the secular terms spoil the approximate result that will
not exhibit the expected periodic behaviour. Grozdanovski and Shepherd\cite
{GS08} applied the well-known Lindstedt-Poincar\'{e} method to remove
secular terms and obtained the first two perturbation corrections to a
two-species system. Consequently, their approximate results exhibit the
expected periodic behaviour. Navarro\cite{N16} also applied the
Lindstedt-Poincar\'{e} method to the same two-species system and obtained
periodic approximate expressions of second order. This author proposed a
symbolic algorithm for the computation of periodic orbits but surprisingly
did not show results of larger order. Navarro and Poveda\cite{NP17} applied
Navarro's approach to a three-species system and derived perturbation
corrections of first and second order for the populations of the three
species for some particular values of the model parameters. All these
studies lead to the conclusion that the Lindstedt-Poincar\'{e} method gives
reasonable results for some model parameters and suggest that the technique
may be useful for the analysis of more realistic and related nonlinear
dynamical problems.

Unfortunately Grozdanovski and Shepherd\cite{GS08} did not explicitly
indicate the initial conditions chosen for the solution of the first-order
differential equations that provide the corrections at every perturbation
order. Since their strategy is not clearly delineated it is difficult to
derive a systematic approach for the calculation of perturbation corrections
of greater order. On the other hand, Navarro\cite{N16} and Navarro and Poveda%
\cite{NP17} put forward a systematic symbolic algorithm but they did not
appear to exploit it to obtain perturbation corrections of large order.
Besides, their presentation of the algorithm appears to be rather obscure
for anybody who is not familiar with such technique.

The aim of this paper is the analysis of the approach proposed by
Grozdanovski and Shepherd\cite{GS08} with the purpose of deriving a
systematic method for the calculation of perturbation corrections of any
order to the Lotka-Volterra model. Such results may give us some clue about
the convergence properties of the perturbation series. In addition to it we
investigate the possibility of generalizing the method for the treatment of
more realistic systems with more than two degrees of freedom.

In section~\ref{sec:L-P} we outline the Lotka-Volterra model and compare
alternative implementations of the Lindstedt-Poincar\'{e} approach. In section~%
\ref{sec:generalization} we put forward a generalization of the method that
enables one to treat all the previously discussed cases. In section~\ref
{sec:Large-order} we carry out a large order calculation of the perturbation
corrections and estimate the radius of convergence of the perturbation
series for the frequency. Finally, in section~\ref{sec:conclusions} we
outline the application of the method to more general and realistic
dynamical systems and draw conclusions.

\section{The Lindstedt-Poincar\'{e} method}

\label{sec:L-P}

In this section we briefly discuss the Lotka-Volterra model and delineate
the application of the Lindsted-Poincar\'{e} technique. For concreteness we
will follow Grozdanovski and Shepherd\cite{GS08} because their treatment of
the dynamical equations is clear and straightforward.

\subsection{The model}

\label{subsec:model}

The dynamical equations for the model are
\begin{equation}
\dot{X}(T)=X(T)\left[ a-bY(T)\right] ,\;\dot{Y}(T)=Y(T)\left[ cX(T)-d\right]
,  \label{eq:dyn-eq-XY}
\end{equation}
where $a$, $b$, $c$ and $d$ are positive parameters and the point indicates
derivative with respect to time. The physical meaning of the parameters is
not relevant for present purposes because we are mainly interested in the
success of the perturbation approach. Besides, most probably nobody will
apply the Lotka-Volterra model to an actual ecological system today because
it is quite unrealistic. The interested reader may resort to the papers
cited above for more information\cite{MSP90,GS08,N16,NP17} (and the
references cited therein).

We can get rid of some model parameters by means of the following
transformations of the independent and dependent variables
\begin{equation}
t=aT,\;x(t)=\frac{c}{d}X(T),\;y(t)=\frac{b}{a}Y(T).
\end{equation}
The resulting equations
\begin{equation}
\dot{x}(t)=x(t)-x(t)y(t),\;\dot{y}(t)=\alpha \left[ -y(t)+x(t)y(t)\right] ,
\label{eq:dyn-eq-xy}
\end{equation}
depend on just one parameter $\alpha =d/a$.

There is a stationary point at $x=1$ and $y=1$. Therefore, if we define
\begin{equation}
x(t)=1+\epsilon \xi (t),\;y(t)=1+\epsilon \eta (t),  \label{eq:x(chi)_y(eta)}
\end{equation}
the resulting dynamical equations will depend on the perturbation parameter $%
\epsilon $
\begin{equation}
\dot{\xi}(t)=-\eta (t)-\epsilon \xi (t)\eta (t),\;\dot{\eta}(t)=\alpha
\left[ \xi (t)+\epsilon \xi (t)\eta (t)\right] .  \label{eq:dyn-eq-xi-eta-1}
\end{equation}
In order to apply the Lindstedt-Poincar\'{e} method we define the
dimensionless time
\begin{equation}
\tau =\omega t,
\end{equation}
where $\omega $ is the unknown frequency of oscillation. In this way we have
\begin{equation}
\omega \dot{\xi}(\tau )=-\eta (\tau )-\epsilon \xi (\tau )\eta (\tau
),\;\omega \dot{\eta}(\tau )=\alpha \left[ \xi (\tau )+\epsilon \xi (\tau
)\eta (\tau )\right] .  \label{eq:dyn-eq-xi-eta-2}
\end{equation}
Following Grozdanovski and Shepherd\cite{GS08} we are using the same symbols
$\xi $ and $\eta $ for the solutions of equations (\ref{eq:dyn-eq-xi-eta-1})
and (\ref{eq:dyn-eq-xi-eta-2}). Besides, we have chosen a dot to indicate
the derivative with respect to either $t$ or $\tau $. Although this practice
may be unwise when one is studying a practical problem and wants to
reconstruct $X(T)$ and $Y(T)$ from $\xi (\tau )$ and $\eta (\tau )$ it is
harmless in the present case because our aim is to show how to obtain
perturbation corrections of large order and study the convergence of the
perturbation series.

\subsection{Perturbation equations}

\label{subsec:PT}

We now assume that $\epsilon $ is a sufficiently small parameter and apply
perturbation theory in the usual way
\begin{eqnarray}
\xi (\tau ) &=&\sum_{j=0}^{\infty }\xi _{j}(\tau )\epsilon ^{j},  \nonumber
\\
\eta (\tau ) &=&\sum_{j=0}^{\infty }\eta _{j}(\tau )\epsilon ^{j},  \nonumber
\\
\omega &=&\sum_{j=0}^{\infty }\omega _{j}\epsilon ^{j}.
\label{eq:epsilon-series}
\end{eqnarray}
From the equations of order zero ($\epsilon =0$) we obtain
\begin{equation}
\xi _{0}(\tau )=A\cos (\tau +\phi ),\,\eta _{0}(\tau )=\sqrt{\alpha }A\sin
(\tau +\phi ),  \label{eq:zero-order-solutions}
\end{equation}
and $\omega _{0}=\sqrt{\alpha }$. On inserting the expansions (\ref
{eq:epsilon-series}) into equations (\ref{eq:dyn-eq-xi-eta-2}) it is not
difficult to show that the perturbation corrections are solutions to
\begin{eqnarray}
\dot{\xi}_{n} &=&-\frac{1}{\sqrt{\alpha }}\eta _{n}+F_{n},\;n=1,2,\ldots ,
\nonumber \\
\dot{\eta}_{n} &=&\sqrt{\alpha }\xi _{n}+G_{n}  \nonumber \\
F_{n} &=&-\frac{1}{\sqrt{\alpha }}\sum_{j=0}^{n-1}\xi _{j}\eta _{n-j-1}-%
\frac{1}{\sqrt{\alpha }}\sum_{j=1}^{n}\omega _{j}\dot{\xi}_{n-j},  \nonumber
\\
G_{n} &=&\sqrt{\alpha }\sum_{j=0}^{n-1}\xi _{j}\eta _{n-j-1}-\frac{1}{\sqrt{%
\alpha }}\sum_{j=1}^{n}\omega _{j}\dot{\eta}_{n-j}.  \label{eq:PT-eqs}
\end{eqnarray}

\subsection{Systematic approach}

\label{subsec:systematic_aproach}

The purpose of this subsection is to derive general expressions for the
solutions to the perturbation equations (\ref{eq:PT-eqs}) that enable us to
develop a systematic algorithm for the calculation of corrections of
sufficiently large order.

Neither Grozdanovski and Shepherd\cite{GS08} nor Navarro\cite{N16} consider
the initial conditions of the perturbation corrections $\xi _{n}(\tau )$ and
$\eta _{n}(\tau )$ explicitly. Here we choose
\begin{equation}
\xi _{n}(0)=0,\;\eta _{n}(0)=0,\;n>0,  \label{eq:bc1}
\end{equation}
because they greatly facilitate the calculation of $A$ and $\phi $ from $%
x(0) $ and $y(0)$:
\begin{equation}
x(0)=1+\epsilon A\cos (\phi ),\;y(0)=1+\epsilon A\sqrt{\alpha }\sin (\phi ).
\label{eq:x(0),y(0)}
\end{equation}
Note that, given $x(0)$ and $y(0)$ we obtain the product $\epsilon A$ and $%
\phi $. Later on we will show why $A$ always appears associated to the
perturbation parameter $\epsilon $ in this particular way.

In order to solve equations (\ref{eq:PT-eqs}) we rewrite them in matrix form
\begin{eqnarray}
\mathbf{\dot{W}}_{n} &=&\mathbf{K\cdot W}_{n}+\mathbf{R}_{n},  \nonumber \\
\mathbf{W}_{n} &=&\left(
\begin{array}{l}
\xi _{n} \\
\eta _{n}
\end{array}
\right) ,\;\mathbf{R}_{n}=\left(
\begin{array}{l}
F_{n} \\
G_{n}
\end{array}
\right) ,  \nonumber \\
\mathbf{K} &=&\frac{1}{\sqrt{\alpha }}\left(
\begin{array}{ll}
0 & -1 \\
\alpha & 0
\end{array}
\right) ,  \label{eq:PT-eqs-mat}
\end{eqnarray}
so that the solution is simply given by
\begin{equation}
\mathbf{W}_{n}(\tau )=\int_{0}^{\tau }\exp \left[ (\tau -s)\mathbf{K}\right]
\cdot \mathbf{R}_{n}(s)\mathbf{\,}ds,  \label{eq:Wn-a}
\end{equation}
where
\begin{equation}
\exp \left( \tau \mathbf{K}\right) =\frac{1}{\sqrt{\alpha }}\left(
\begin{array}{ll}
\sqrt{\alpha }\cos (\tau ) & -\sin (\tau ) \\
\alpha \sin (\tau ) & \sqrt{\alpha }\cos (\tau )
\end{array}
\right) .  \label{eq:exp_mat}
\end{equation}
Note that equation (\ref{eq:Wn-a}) is consistent with the initial conditions
(\ref{eq:bc1}).

In order to identify the resonant terms that would give rise to
secular terms we rewrite the first-order differential equations as
second order ones; for example
\begin{equation}
\ddot{\xi}_{n}=-\xi _{n}+\dot{F}_{n}-\frac{1}{\sqrt{\alpha }}G_{n}.
\end{equation}
Therefore, we set $\omega _{n}$ so that
\begin{eqnarray}
\int_{0}^{2\pi }\left[ \dot{F}_{n}(\tau )-\frac{1}{\sqrt{\alpha }}G_{n}(\tau
)\right] \sin (\tau +\phi )\,d\tau &=&0,\,  \nonumber \\
\int_{0}^{2\pi }\left[ \dot{F}_{n}(\tau )-\frac{1}{\sqrt{\alpha }}G_{n}(\tau
)\right] \cos (\tau +\phi )\,d\tau &=&0.  \label{eq:secular_remove}
\end{eqnarray}
These equations are a generalization of the Lemma 1 in the paper by
Grozdanovski and Shepherd\cite{GS08} and the proposal of Navarro\cite{N16}.
It is worth noting that the same value of $\omega _{n}$ satisfies both
equations (\ref{eq:secular_remove}). We are not aware of a rigorous proof of
this result but we can test it by means of our calculations of large order.
This point was not discussed in the earlier papers on the application of the
Lindstedt-Poincar\'{e} method to multidimensional systems\cite{GS08,N16,NP17}
probably because they did not try to obtain a set of explicit equations for
a systematic application of the approach.

Unfortunately, the perturbation corrections obtained in this way are
considerably more complicated than those derived by Grozdanovski and Shepherd%
\cite{GS08}. For example, at first order we obtain
\begin{eqnarray}
\xi _{1}(\tau ) &=&A^{2}\left[ \frac{\sin {\left( \phi \right) }}{4}-\frac{%
\sqrt{\alpha }\cos {\left( 3\phi \right) }}{12}-\frac{\sin {\left( 3\phi
\right) }}{12}-\frac{\sqrt{\alpha }\cos {\left( \phi \right) }}{4}\right]
\sin (\tau +\phi )  \nonumber \\
&&+\frac{A^{2}\sqrt{\alpha }}{6}\sin [2(\tau +2\phi )]  \nonumber \\
&&+A^{2}\left[ \frac{\sqrt{\alpha }\sin {\left( 3\phi \right) }}{12}-\frac{%
\cos {\left( \phi \right) }}{4}-\frac{\cos {\left( 3\phi \right) }}{12}-%
\frac{\sqrt{\alpha }\sin {\left( \phi \right) }}{4}\right] \cos (\tau +\phi )
\nonumber \\
&&+\frac{A^{2}}{3}\cos [2(\tau +\phi )],  \nonumber \\
\eta _{1}(\tau ) &=&A^{2}\left[ \frac{\alpha \sin {\left( 3\phi \right) }}{12%
}-\frac{\sqrt{\alpha }\cos {\left( 3\phi \right) }}{12}-\frac{\sqrt{\alpha }%
\cos {\left( \phi \right) }}{4}-\frac{\alpha \sin {\left( \phi \right) }}{4}%
\right] \sin (\tau +\phi )  \nonumber \\
&&+\frac{A^{2}\sqrt{\alpha }}{6}\sin [2(\tau +\phi )]  \nonumber \\
&&+A^{2}\left[ \frac{\alpha \cos {\left( 3\phi \right) }}{12}+\frac{\sqrt{%
\alpha }\sin {\left( 3\phi \right) }}{12}+\frac{\alpha \cos {\left( \phi
\right) }}{4}-\frac{\sqrt{\alpha }\sin {\left( \phi \right) }}{4}\right]
\cos (\tau +\phi )  \nonumber \\
&&-\frac{A^{2}\alpha }{3}\cos [2(\tau +\phi )].  \label{eq:xi1_eta1_my}
\end{eqnarray}
The coefficients of $\sin [2(\tau +2\phi )]$ and $\cos [2(\tau +2\phi )]$
agree with the ones derived earlier by those authors and the remaining terms
are necessary to satisfy the initial conditions (\ref{eq:bc1}). We also find
that $\omega _{1}=0$ removes the resonant terms. The perturbation
corrections derived by Navarro\cite{N16} with somewhat different initial
conditions appear to be simpler but they are restricted to $\alpha =1$.

The perturbation corrections of second order are so complicated that we do
not show them here. Besides, $\omega _{3}$ is nonzero and a rather
cumbersome function of $\alpha $ and $\phi $:
\begin{eqnarray}
\omega _{3} &=&\frac{A^{3}\sqrt{\alpha }\left( \alpha +1\right) \cos {\left(
3\phi \right) }}{144}-\frac{A^{3}\alpha \left( \alpha +1\right) \sin {\left(
3\phi \right) }}{144}  \nonumber \\
&&+\frac{A^{3}\sqrt{\alpha }\left( \alpha +1\right) \cos {\left( \phi
\right) }}{48}+\frac{A^{3}\alpha \left( \alpha +1\right) \sin {\left( \phi
\right) }}{48}.  \label{eq:omega_3}
\end{eqnarray}
The occurrence of rather too complicated perturbation corrections appears to
be the price that one has to pay for obtaining the simpler equations (\ref
{eq:x(0),y(0)}) for the calculation of $\epsilon A$ and $\phi $.

\begin{figure}[tbp]
\begin{center}
\includegraphics[width=9cm]{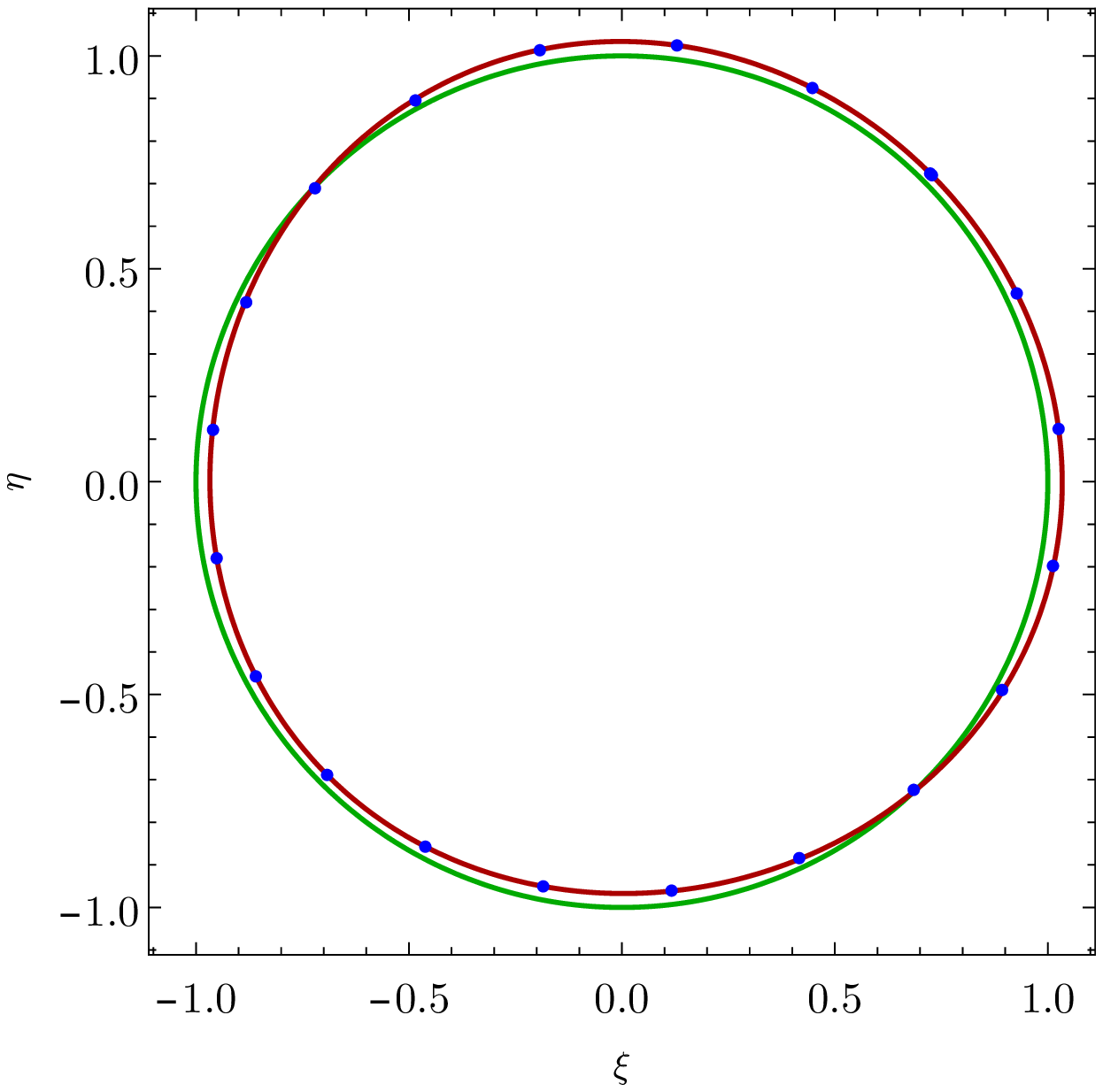}
\end{center}
\caption{Curve $\eta(\xi)$ calculated by means of perturbation theory of
order zero (dashed, green line), up to second order (continuous, red line)
and numerically (blue points) for $\alpha = 1$, $\epsilon A=0.1$ and $\phi =
\pi/4$}
\label{fig:eta(xi)}
\end{figure}

At first sight the dependence of $\omega _{n}$ on the phase $\phi $ may
appear to be the consequence of a wrong calculation. However, we have
verified that present solutions already satisfy the perturbation equations
and comparison with numerical results reveals a good agreement. For example,
figure~\ref{fig:eta(xi)} compares the curve $\eta (\xi )$ for $\alpha =1$, $%
\epsilon A=0.1$ and $\phi =\pi /4$ calculated by perturbation theory of
zeroth and second order and an accurate numerical result. We appreciate that
the addition of the perturbation corrections shown above already improves
the analytical results. In the next subsection we will explain the reason
for the discrepancy between our expressions and those of Grozdanovski and
Shepherd\cite{GS08} in a more transparent way.

\subsection{The straightforward Fourier expansion}

\label{subsec:fourier}

The purpose of this subsection is merely to show why it is possible to
obtain many different solutions at every order of perturbation theory. We
may solve the differential perturbation equations (\ref{eq:PT-eqs}) by
inserting Fourier expansions of the form
\begin{eqnarray}
\xi _{n}(\tau ) &=&\sum_{j=1}^{n+1}a_{1j}^{(n)}\sin [j(\tau +\phi
)]+\sum_{j=0}^{n+1}b_{1j}^{(n)}\cos [j(\tau +\phi )],  \nonumber \\
\eta _{n}(\tau ) &=&\sum_{j=1}^{n+1}a_{2j}^{(n)}\sin [j(\tau +\phi
)]+\sum_{j=0}^{n+1}b_{2j}^{(n)}\cos [j(\tau +\phi )].
\label{eq:Fourier_expansion}
\end{eqnarray}
For the first order we obtain (we omit the superscript for simplicity)
\begin{eqnarray}
a_{11}+\frac{b_{21}}{\sqrt{\alpha }} &=&0,\;b_{11}-\frac{a_{21}}{\sqrt{%
\alpha }}=0,  \nonumber \\
a_{12} &=&\frac{A^{2}\sqrt{\alpha }}{6},\;b_{12}=\frac{A^{2}}{3},\;a_{22}=%
\frac{A^{2}\sqrt{\alpha }}{6},\;b_{22}=-\frac{A^{2}\alpha }{3}.
\label{eq:aij_first_order}
\end{eqnarray}
We appreciate that if we choose $a_{11}=b_{21}=b_{11}=a_{21}=0$ we obtain
exactly the results of Grozdanovski and Shepherd\cite{GS08}. However, there
is an infinite number of perfectly valid solutions that emerge from
arbitrary choices of $a_{11}$, $b_{21}$, $b_{11}$ and $a_{21}$ provided that
they satisfy the ratios $\frac{b_{21}}{a_{11}}=-\frac{a_{21}}{b_{11}}=-\sqrt{%
\alpha }$. One of such possible solutions is that shown above that satisfies
the boundary conditions (\ref{eq:bc1}). The solutions derived by the authors
just mentioned seem to be the simplest ones and are therefore most
convenient for large-order calculations. We should find suitable general
conditions to produce such simple results at every order of perturbation
theory.

\section{Generalization of the systematic approach}

\label{sec:generalization}

In the preceding sections we discussed two possible solutions: those that
lead to the simple initial conditions (\ref{eq:x(0),y(0)}) and those that
are considerably simpler but lead to somewhat complicated initial
conditions. The problem at hand is that we have not yet specified the
initial conditions for the perturbation equations that lead to the latter.
Simpler solutions are obviously most convenient for the calculation of
analytic perturbation corrections of large order because they will render
the computation algorithm more efficient and less time and memory consuming.

Fortunately, it is not difficult to make the general approach of subsection~%
\ref{subsec:systematic_aproach} more flexible so that it yields results that
are as simple as those of Grozdanovski and Shepherd\cite{GS08}. We simply
choose general initial conditions of the form
\begin{equation}
\xi _{n}(0)=a_{n},\;\eta _{n}(0)=b_{n}.  \label{eq:bc2}
\end{equation}
Now the solution to the matrix perturbation equations (\ref{eq:PT-eqs-mat})
is given by
\begin{equation}
\mathbf{W}_{n}(\tau )=\exp \left( \tau \mathbf{K}\right) \cdot \left(
\begin{array}{l}
a_{n} \\
b_{n}
\end{array}
\right) +\int_{0}^{\tau }\exp (\left[ \tau -s)\mathbf{K}\right] \cdot
\mathbf{R}_{n}(s)\mathbf{\,}ds,  \label{eq:Wn-b}
\end{equation}
and we can choose the arbitrary real numbers $a_{n}$ and $b_{n}$ so that the
pair of solutions at order $n$ is as simple as possible. In what follows we
simply set them so that the coefficients of $\sin (\tau +\phi )$ and $\cos
(\tau +\phi )$ in $\xi _{n}(\tau )$ vanish (we can, of course, choose $\eta
_{n}(\tau )$ instead). More precisely, $a_{n}$ and $b_{n}$ are solutions to
the equations
\begin{equation}
\int_{0}^{2\pi }\xi _{n}(\tau )\sin (\tau +\phi )\,d\tau =0,\;\int_{0}^{2\pi
}\xi _{n}(\tau )\cos (\tau +\phi )\,d\tau =0.  \label{eq:eqs-for-an_bn}
\end{equation}
It is obvious that in this way the solutions $\xi _{n}(\tau )$ and $\eta
_{n}(\tau )$ are completely determined.

To first order we obtain
\begin{eqnarray}
a_{1} &=&A^{2}\left[ \frac{\sqrt{\alpha }\sin {\left( 2\phi \right) }}{6}-%
\frac{\alpha \cos {\left( 2\phi \right) }}{3}\right] ,  \nonumber \\
b_{1} &=&A^{2}\left[ \frac{\sqrt{\alpha }\sin {\left( 2\phi \right) }}{6}-%
\frac{\alpha \cos {\left( 2\phi \right) }}{3}\right] ,
\end{eqnarray}
consistent with the results of Grozdanovski and Shepherd\cite{GS08} for $\xi
_{1}(\tau )$ and $\eta _{1}(\tau )$.

To second order we have
\begin{eqnarray}
a_{2} &=&A^{3}\left[ \frac{\sqrt{\alpha }\cos {\left( 2\phi \right) }}{%
16\sin {\left( \phi \right) }}+\frac{\left( 3-\alpha \right) \cos {\left(
3\phi \right) }}{32}-\frac{\sqrt{\alpha }\cos {\left( 4\phi \right) }}{%
16\sin {\left( \phi \right) }}\right] ,  \nonumber \\
b_{2} &=&A^{3}\left[ \frac{\alpha \cos {\left( \phi \right) }}{12}+\frac{%
\sqrt{\alpha }\left( 1-\alpha \right) \sin {\left( \phi \right) }}{24}-\frac{%
\alpha \cos {\left( 3\phi \right) }}{8}\right.  \nonumber \\
&&\left. +\frac{\sqrt{\alpha }\left( 1-3\alpha \right) \sin {\left( 3\phi
\right) }}{32}\right] ,
\end{eqnarray}
and
\begin{eqnarray}
\xi _{2} &=&A^{3}\left\{ \frac{\sqrt{\alpha }}{8}\sin [2(\tau +\phi )]+\frac{%
\left( 3-\alpha \right) }{32}\cos [2(\tau +\phi )]\right\} ,  \nonumber \\
\eta _{2} &=&A^{3}\left\{ \frac{\sqrt{\alpha }\left( 1-\alpha \right) }{24}%
\sin (\tau +\phi )+\frac{\sqrt{\alpha }\left( 1-3\alpha \right) }{32}\sin
[2(\tau +\phi )]\right.  \nonumber \\
&&\left. +\frac{\alpha }{12}\cos (\tau +\phi )-\frac{\alpha }{8}\cos [2(\tau
+\phi )]\right\} .
\end{eqnarray}
These solutions are different from those of Grozdanovski and Shepherd\cite
{GS08} but all of them satisfy the perturbation equations. We would have
obtained exactly their results if we had chosen $a_{2}$ and $b_{2}$ that
make the coefficients of $\sin (\tau +\phi )$ and $\cos (\tau +\phi )$ in $%
\eta _{2}(\tau )$ vanish. We just did it in this way to stress the ambiguity
of the results already outlined above in subsection~\ref{subsec:fourier}.
Note that if we substitute $\eta _{n}(\tau )$ for $\xi _{n}(\tau )$ in
equations (\ref{eq:eqs-for-an_bn}) we modify the, in principle arbitrary,
initial conditions for the solutions to the perturbation equations (\ref
{eq:PT-eqs}). In either case we have $\omega _{3}=0$.

The perturbation corrections of third order are given by the coefficients
(we again omit the superscripts)
\begin{eqnarray}
a_{12} &=&\frac{A^{4}\sqrt{\alpha }\left( \alpha -11\right) }{864},\;a_{14}=%
\frac{A^{4}\sqrt{\alpha }\left( 125-13\alpha \right) }{2160},  \nonumber \\
b_{12} &=&\frac{A^{4}\left( \alpha +7\right) }{432},\;b_{14}=\frac{%
A^{4}\left( 13-20\alpha \right) }{540},  \nonumber \\
a_{22} &=&\frac{A^{4}\sqrt{\alpha }\left( 25\alpha +13\right) }{864}%
,\;a_{24}=\frac{A^{4}\sqrt{\alpha }\left( 13-125\alpha \right) }{2160},
\nonumber \\
b_{22} &=&\frac{A^{4}\alpha \left( 5\alpha -1\right) }{432},\;b_{24}=\frac{%
A^{4}\alpha \left( 13\alpha -20\right) }{540}.
\end{eqnarray}
From all these results we obtain
\begin{equation}
\omega _{4}=-\frac{A^{4}\sqrt{\alpha }\left( 5\alpha ^{2}+34\alpha
+29\right) }{6912},
\end{equation}
that was not calculated by earlier authors as far as we know.

Assisted by available computer algebra software we have calculated $\xi _{1}$%
, $\xi _{2}$, $\ldots $, $\xi _{7}$, $\eta _{1}$, $\eta _{2}$, $\ldots $, $%
\eta _{7}$ interactively and our analytical results suggest that $\omega
_{2n+1}=0$, $n=0,1,\ldots $ for the boundary conditions (\ref{eq:bc2}) given
by (\ref{eq:eqs-for-an_bn}). Here we just show the next two perturbation
corrections to the frequency:
\begin{eqnarray}
\omega _{6} &=&\frac{A^{6}\sqrt{\alpha }\left( 97\alpha ^{3}-645\alpha
^{2}-2925\alpha -2183\right) }{3317760},  \nonumber \\
\omega _{8} &=&\frac{A^{8}\sqrt{\alpha }\left( 102293\alpha
^{4}+188228\alpha ^{3}-763890\alpha ^{2}-2581852\alpha -1732027\right) }{%
14332723200}.  \nonumber \\
&&
\end{eqnarray}
We want to point out that up to this point we have carried out the
calculation order by order interactively (that is to say: without
programming the equations for the calculation of the perturbation
corrections). Obviously, this strategy is unsuitable for the calculation of
large order we are interested in. However, even in this rather inefficient
way we derived perturbation corrections of order larger than those shown by
Navarro\cite{N16} who proposed a symbolic algorithm for this purpose.

In closing this section we want to make a couple of considerations about the
perturbative solution of this model. To begin with note that we can rewrite
equation (\ref{eq:dyn-eq-xi-eta-2}) as
\begin{eqnarray}
\omega \frac{d}{d\tau }\left( \frac{\xi }{A}\right) &=&-\frac{\eta }{A}%
-A\epsilon \frac{\xi }{A}\frac{\eta }{A},\;  \nonumber \\
\omega \frac{d}{d\tau }\left( \frac{\eta }{A}\right) &=&\alpha \left[ \frac{%
\xi }{A}+A\epsilon \frac{\xi }{A}\frac{\eta }{A}\right] .
\end{eqnarray}
Therefore, we can obtain $\xi (\tau ,\epsilon ,A)=A\xi (\tau ,A\epsilon ,1)$
and $\eta (\tau ,\epsilon ,A)=A\eta (\tau ,A\epsilon ,1)$ from the solutions
to the perturbation equations for $A=1$ and perturbation parameter $%
a=\epsilon A$. This transformation is convenient because we will not have $A$
in the analytic solutions to the perturbation corrections which results in
the use of less computer memory. Note that Grozdanovski and Shepherd\cite
{GS08} also defined the parameter $a$ to write their expressions for $x(t)$
and $y(t)$ in a more compact way. However, they did not appear to exploit
this fact in a systematic way.

\section{Large-order calculations}

\label{sec:Large-order}

In this section we will show that the algorithm discussed in section~\ref
{sec:generalization} is actually useful for the calculation of perturbation
corrections of sufficiently large order and will exploit the fact that the
perturbation equations (\ref{eq:PT-eqs-mat}) and (\ref{eq:Wn-b}),
supplemented by (\ref{eq:secular_remove}) and (\ref{eq:eqs-for-an_bn}), are
suitable for programming in available computer algebra systems. For
concreteness and simplicity we will focus on the perturbation series for the
frequency
\begin{equation}
\omega =1+\sum_{j=1}^{\infty }c_{j}(\alpha )a^{2j},\;c_{j}=\omega
_{2j},\;a=\epsilon A,  \label{eq:omega_series}
\end{equation}
and will try to determine its radius of convergence.

In general, the radius of convergence $r_{c}$ of the power-series expansion
\begin{equation}
f(z)=\sum_{j=0}^{\infty }c_{j}z^{j},
\end{equation}
is determined by the singularity $z_{s}$ of the function $f(z)$
closest to the origin: $r_{c}=|z_{s}|$. There are many ways of
estimating the singularities of an unknown function from its known
power-series expansion. One of them is given by the Pad\'{e}
approximants\cite{S74}
\begin{eqnarray}
f[K,L,z] &=&\frac{P_{K}(z)}{Q_{L}(z)},  \nonumber \\
P_{K}(z) &=&\sum_{j=0}^{K}p_{j}z^{j},  \nonumber \\
Q_{L}(z) &=&\sum_{j=0}^{L}q_{j}z^{j},  \label{eq:Padé}
\end{eqnarray}
where the coefficients $p_{j}$ and $q_{k}$ are chosen so that
\begin{equation}
f[K,L,z]=\sum_{j=0}^{K+L+1}c_{j}z^{j}+O\left( z^{K+L+2}\right) .
\end{equation}
It is commonly assumed that the stable zero of $Q(z)$ (as $K$ and
$L$ increases) closest to the origin provides an estimate of
$z_{s}$.

In some cases it is more convenient to resort to quadratic Hermite-Pad\'{e}
approximants\cite{S74}
\begin{eqnarray}
&&P_{K}(z)\left( f[K,L,M,z]\right) ^{2}+Q_{L}(z)f[K,L,M,z]+R_{M}(z),
\nonumber \\
P_{K}(z) &=&\sum_{j=0}^{K}p_{j}z^{j},  \nonumber \\
Q_{L}(z) &=&\sum_{j=0}^{L}q_{j}z^{j},  \nonumber \\
R_{M}(z) &=&\sum_{j=0}^{M}r_{j}z^{j},  \label{eq:Hermite-Padé}
\end{eqnarray}
where the coefficients $p_{j}$, $q_{k}$ and $r_{m}$ are chosen so that
\begin{equation}
f[K,L,M,z]=\sum_{j=0}^{K+L+M+1}c_{j}z^{j}+O\left( z^{K+L+M+2}\right) .
\end{equation}
In this case the singularity closest to the origin is a stable root of
\begin{equation}
Q_{L}(z)^{2}-4P_{K}(z)R_{M}(z)=0.
\end{equation}

\begin{figure}[tbp]
\begin{center}
\includegraphics[width=9cm]{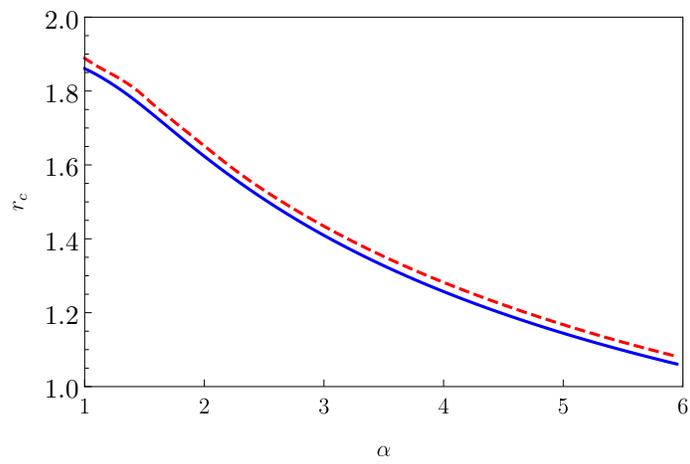}
\end{center}
\caption{Radius of convergence of the series for the frequency as function
of $\alpha$. The dashed and continuous lines are results from Pad\'e and
Hermite-Pad\'e approximants.}
\label{fig:rc(alpha)}
\end{figure}

With the algorithm of section~\ref{sec:generalization} we have been able to
obtain $c_{j}=\omega _{2j}$ for $j=1,2,\ldots ,22$ as analytical functions
of $\alpha $. By means of diagonal Pad\'{e} ($K=L$) and Hermite-Pad\'{e} ($%
K=L=M$) approximants we estimated $r_{c}(\alpha )$ for the expansion
variable $z=a^{2}$. Figure~\ref{fig:rc(alpha)} shows a good agreement
between both types of approximants. We appreciate that the radius of
convergence is a monotonously decreasing function of the model parameter $%
\alpha $. In a recent paper Amore et al\cite{ABF17} calculated the radius of
convergence of the frequency of the van der Pol oscillator with
unprecedented accuracy by means of Hermite-Pad\'{e} approximants constructed
from the Lindstedt-Poincar\'{e} series with an extremely large number of
terms. We are therefore confident of the accuracy of present results.

As an additional verification of the accuracy of our results we have carried
out a perturbation calculation of order $62$ for $\alpha =1$ and obtained $%
r_{c}=3.462532$ and $r_{c}=3.457033$ with Hermite-Pad\'{e} approximants $%
f[7,7,7,z]$ and $f[10,10,10,z]$, respectively. On the other hand, the
diagonal Pad\'{e} approximants exhibit a stable pole close to the origin at $%
z=3.5$. Based on these results we can safely conclude that $r_{c}(1)\approx
3.46$.

As mentioned before Navarro\cite{N16} developed a symbolic algorithm for the
computation of the Lindstedt-Poincar\'{e} perturbation corrections and
applied it to the Lotka-Volterra model but did not show any results beyond
second order. As far as we know there is no perturbation calculation in the
literature of order as high as the one shown here. The usefulness of such
calculation is obvious because it enables us to estimate the practical range
of utility of the Lindstedt-Poincar\'{e} perturbation theory for the
treatment of dynamical systems. In the case of the Lotka-Volterra model we
clearly appreciate that this approximation is not valid if the initial
populations $x(0)$ and $y(0)$ are such that $|\epsilon A|>r_{c}(\alpha )$.

\section{Further comments and conclusions}

\label{sec:conclusions}

In this paper we have developed a systematic method for the
application of the Linstedt-Poincar\'{e} perturbation theory to
the Lotka-Volterra model. In particular we discussed the initial
conditions for the perturbation equations that were not taken into
account explicitly in earlier papers\cite {GS08,N16}. Present
analysis reveals that one can obtain an infinite number of
solutions to the perturbation equations and the choice of one of
them depends solely on convenience. Here we weighted the
possibility of simpler expressions for the calculation of the
parameters $\epsilon A$ and $\phi $ on the one side against the
simplicity of the solutions on the other. In the latter case we
obtained perturbation corrections of considerably larger order
than those derived earlier\cite{GS08,N16}. From them we could
estimate the radius of convergence of the perturbation series for
the frequency. This result is important for the estimation of the
range of validity of the approximate perturbation solutions to the
dynamical equations. It shows that the resulting analytical
expressions are bounded to fail for some initial conditions.

Present approach can be easily generalized to periodic nonlinear systems of
any number of dynamical variables. For example, if we can rewrite the
perturbation corrections to the dynamical equations in the form
\begin{equation}
\mathbf{\dot{W}}_{n}=\mathbf{K}\cdot \mathbf{W}_{n}+\mathbf{R}_{n},
\end{equation}
where $\mathbf{K}$ is an $N\times N$ matrix and $\mathbf{W}_{n}$ and $%
\mathbf{R}_{n}$ are $N-$dimensional column vectors, then the
solution to the perturbation equation of order $n$ is given by
\begin{equation}
\mathbf{W}_{n}=\exp \left( \tau \mathbf{K}\right) \mathbf{V}%
_{n}+\int_{0}^{\tau }\exp \left[ (\tau -s)\mathbf{K}\right] \mathbf{R}%
_{n}(s)\,ds,
\end{equation}
where $\mathbf{V}_{n}$ is an $N-$dimensional column vector with arbitrary
elements that we choose in order to obtain the simplest solutions. In order
to carry out this calculation we just need $\exp (\tau \mathbf{K})$ but its
construction is a textbook excercise\cite{A69}.

\section*{Acknowledgments}

The research of P. Amore was supported by the Sistema Nacional de
Investigadores (M\'exico)

\end{document}